\documentclass[preprint,prd,nofootinbib,tightenlines,amsmath]{revtex4}
\usepackage{graphicx}

\begin{document}
\baselineskip=14pt \parskip=3pt

\vspace*{3em}

\title{Hints of Standard Model Higgs Boson at the LHC \\ and Light Dark Matter Searches}

\author{Xiao-Gang He$^{1,2,3}$}
\author{Bo Ren$^1$}
\author{Jusak Tandean$^{4}$}
\affiliation{$^1$INPAC, Department of Physics,
Shanghai Jiao Tong University, Shanghai, China \vspace*{1ex}
\\
$^2$Department of Physics and Center for Theoretical Sciences,
National Taiwan University, Taipei 106, Taiwan \vspace*{1ex}
\\
$^3$\mbox{Department of Physics, National Tsing Hua University}, \\ and National
Center for Theoretical Sciences, Hsinchu 300, Taiwan \vspace*{1ex}
\\
$^4$Department of Physics and Center~for~Mathematics and Theoretical Physics,
National Central University, Chungli 320, Taiwan \vspace*{1ex}
\\
}



\begin{abstract}
The most recent results of searches at the LHC for the Higgs boson $h$ have turned up
possible hints of such a particle with mass $m_h^{}$ about 125\,GeV consistent with
standard model (SM) expectations.
This has many potential implications for the SM and beyond.
We consider some of them in the contexts of a simple Higgs-portal dark matter (DM) model,
the SM plus a real gauge-singlet scalar field $D$ as the DM candidate,
and a couple of its variations.
In the simplest model with one Higgs doublet and three or four generations of fermions,
for $D$ mass \,$m_D^{}<m_h^{}/2$\, the invisible decay \,$h\to DD$\, tends to have a substantial
branching ratio.
If future LHC data confirm the preliminary Higgs indications, $m_D^{}$ will have to
exceed~$m_h^{}/2$.  To keep the DM lighter than~$m_h^{}/2$, one will need to extend
the model and also satisfy constraints from DM direct searches.
The latter can be accommodated if the model provides sizable isospin violation in
the DM-nucleon interactions.
We explore this in a two-Higgs-doublet model combined with the scalar field~$D$.
This  model can offer a 125-GeV SM-like Higgs and a~light DM candidate having isospin-violating
interactions with nucleons at roughly the required level, albeit with some degree of fine-tuning.
\end{abstract}

\maketitle

\section{Introduction}

The latest searches for the standard model (SM) Higgs boson performed by the ATLAS and CMS
Collaborations at the LHC have come up with tantalizing hints of the particle~\cite{lhc}.
They observed modest excesses of events compatible with a SM Higgs $h$ having mass $m_h^{}$ in
the range from 124 to 126~GeV, but the statistical significance of the excesses is not enough
for making any conclusive statement on the Higgs existence or nonexistence~\cite{lhc}.
Interestingly, these numbers are consistent with \,$m_h^{}=125_{-10}^{+8}$~GeV\, from
the SM complete fit to electroweak precision data plus constraints from direct Higgs searches
at LEP and Tevatron~\cite{gfitter1}.
Needless to say, if upcoming measurements confirm these preliminary findings
at the LHC to be glimpses of the SM Higgs, or a SM-like Higgs,
the implications will be far-reaching for efforts to extend the SM, as physics beyond the SM
is still necessary to account for, among other things, the observed evidence for dark matter.
Particularly, all new-physics models would have to include such a~spinless boson as one of
their ingredients.

Simultaneously with the quest for the Higgs, a number of underground experiments have for
years been searching directly for weakly interacting massive particle~(WIMP) dark
matter~(DM) by looking for the recoil energy of nuclei caused by WIMPs colliding with nucleons.
Intriguingly, some of these searches have turned up excess events which may have been WIMP
signals.
Specifically, the DAMA, CoGeNT, and CRESST-II
Collaborations~\cite{Bernabei:2010mq,Aalseth:2010vx,cresst} have acquired data that seem
to be pointing to light WIMPs of mass in the region roughly from~5~to~30~GeV and
spin-independent WIMP-nucleon scattering cross-sections of order \,$10^{-42}$ to
$10^{-40}$~cm$^2$,\, although the respective ranges preferred by these experiments do not
fully agree with each other.
In contrast, other direct searches, especially by the CDMS, XENON, and SIMPLE
Collaborations~\cite{Akerib:2010pv,Angle:2011th,Aprile:2011hi,Felizardo:2011uw}, still have
not produced any WIMP evidence.
Although presently for WIMP masses under 15\,GeV the latter null results are still
controversial~\cite{Hooper:2010uy} and future WIMP searches with improved sensitivity may
eventually settle this issue definitively, there may be alternative explanations
worth exploring which can reconcile these disagreeing findings on~DM.

Since the various DM searches employed different target materials for WIMP detection,
one of the possible resolutions to the light-DM controversy that have been proposed is
to allow large isospin violation in the WIMP-nucleon
interactions~\cite{Bernabei:2001ve,Kurylov:2003ra,Feng:2011vu}.
It turns out that the tension between the conflicting direct-search results can be partially
eased if the effective WIMP couplings $f_p^{}$ and $f_n^{}$ to the proton and neutron,
respectively, satisfy the
ratio~\,$f_n^{}/f_p^{}\simeq-0.7$\,~\cite{Kurylov:2003ra,Feng:2011vu,Kopp:2011yr}.

In view of these developments in the hunts for the Higgs and for~DM, it is of interest
to consider some of their implications within the context of a relatively simple framework.
For if the two sectors are intimately connected, detecting the signs of one of them could
shine light on still hidden elements of the other.

The most economical model possessing both a Higgs boson and a WIMP candidate is
the~SM+D, which is the SM expanded with the addition of a real gauge-singlet scalar field $D$
dubbed darkon acting as
the~WIMP~\cite{Silveira:1985rk,Burgess:2000yq,He:2008qm,He:2010nt,He:2011de}.
This model predicts that the Higgs decay for \,$m_h^{}\sim120$-130\,GeV\, is
highly dominated by the invisible mode \,$h\to DD$\,
if the darkon mass~\,$m_D^{}<m_h^{}/2$,\, except when $m_D^{}$ is close to, but less
than,~$m_h^{}/2$\,~\cite{He:2008qm,He:2010nt,He:2011de}.
Thus, if a~Higgs with \,$m_h^{}\sim125$\,GeV\,~and characteristics within the SM preference
manifests itself unambiguously in LHC data, the parameter space of the SM+D with a light
darkon will be strongly diminished.
On the other hand, for \,$m_D^{}\ge m_h^{}/2$\, the model is consistent with the existence of
such a~Higgs, as its decay pattern is unaffected  at leading order by the presence of the darkon,
although limited portions of this $m_D^{}$ zone are already excluded by direct-search data.

In order to have a Higgs compatible with the LHC indications as well as a light-WIMP candidate,
one must therefore expand the~SM+D.
This motivates us to study in this paper a~slightly extended model we call THDM+D, which is
a two-Higgs-doublet model combined with a darkon.\footnote{\baselineskip=11pt%
Various aspects of the THDM+D
were previously addressed in~Refs.\,\cite{He:2008qm,Bird:2006jd,He:2007tt,Cai:2011kb}.}
It can offer such a~Higgs and an ample amount of viable parameter space for the darkon.
The model can also supply isospin-violating WIMP-nucleon interactions at about the desired level,
although this will require fine-tuning to some extent.
In addition, the potential presence of tree-level flavor-changing couplings of the neutral
Higgs bosons in the THDM+D implies that the Higgs-mediated top-quark decays \,$t\to(u,c)DD$,\,
if kinematically allowed, would contribute to the decays \,$t\to(u,c)$ plus missing energy.
If these decays have rates that are sufficiently amplified to be measurable at the~LHC,
they can be another avenue to probe the darkon~\cite{He:2007tt}.

In the next section, we consider in more detail some implications of the possible discovery of
a~Higgs having SM-like properties for the SM+D with either three or four sequential
generations of fermions (hereafter referred to as SM3+D or SM4+D, respectively).
In~Sec.\,\ref{thdm+d} we describe the THDM+D with a SM-like Higgs and study its prediction
for WIMP-nucleon cross-sections in the limit that isospin violation is negligible in
the WIMP-nucleon effective couplings.
In~Sec.\,\ref{ivdm} we treat the THDM+D case with isospin-violating WIMP-nucleon interactions.
We give our conclusions in~Sec.\,\ref{concl}.

\section{Standard model plus darkon\label{sm+d}}

In the DM sector of the SM+D, to ensure the stability of the darkon $D$ as the~DM, one assumes
it to be a singlet under the gauge groups of the model and introduces a~$Z_2$ symmetry under
which \,$D\to-D$,\, all the other fields being unaffected.
Requiring that the darkon Lagrangian be renormalizable then implies that it has
the form~\cite{Silveira:1985rk,Burgess:2000yq}
\begin{eqnarray}  \label{DH}
{\cal L}_D^{} \,=\, \mbox{$\frac{1}{2}$}\partial^\mu D\,\partial_\mu^{}D
-\mbox{$\frac{1}{4}$}\lambda_D^{}D^4-\mbox{$\frac{1}{2}$}m_0^2D^2-\lambda D^2 H^\dagger H ~, ~
\end{eqnarray}
where $\lambda_D^{}$,  $m_0^{}$, and $\lambda$  are free parameters and $H$ is the Higgs
doublet containing the physical Higgs field~$h$, in the notation of Ref.\,\cite{He:2010nt}
which has additional details on the model.
Clearly its darkon sector has very few free parameters, only two of which, besides
$m_h^{}$, are relevant here: the Higgs-darkon coupling~$\lambda$, which determines the darkon
relic density, and the darkon mass~\,$m_D^{}=(m^2_0+\lambda v^2)^{1/2}$,\,
where \,$v=246$\,GeV\, is the Higgs vacuum expectation value~(VEV).

As explained recently in Ref.\,\cite{He:2011de}, constraints on the darkon in the SM+D
(either SM3+D or~SM4+D) from a number of rare meson decays with missing energy and from
DM direct searches, pending a definitive resolution to the light-WIMP puzzle, allow
the darkon mass values \,$2.5{\rm\,GeV}\le m_D^{}\le15$\,GeV\, consistent with the light-WIMP
hypothesis and those not far from, but smaller than,~$m_h^{}/2$.
For such masses, the invisible decay mode \,$h\to D D$\, can dominate the Higgs total width
depending on~$m_h^{}$, making the Higgs still hidden from detection, and as a~consequence
significant portions of the $m_h^{}$ ranges in the SM3~(SM4) excluded by the current LHC data
may be made viable again in the~SM3+D~(SM4+D).\footnote{\baselineskip=11pt%
Here, as in Refs.\,\cite{He:2010nt,He:2011de}, all the fourth-generation fermions in the SM4+D
are assumed to be unstable. If the fourth-generation neutrino is stable, it can be a DM candidate
which may also contribute to the Higgs invisible decay and makes up a fraction of the DM relic
density, as was studied in Ref.\,\cite{Belotsky:2002ym} in the absence of the darkon.}

\begin{figure}[b]
\, \, \, \includegraphics[width=80mm]{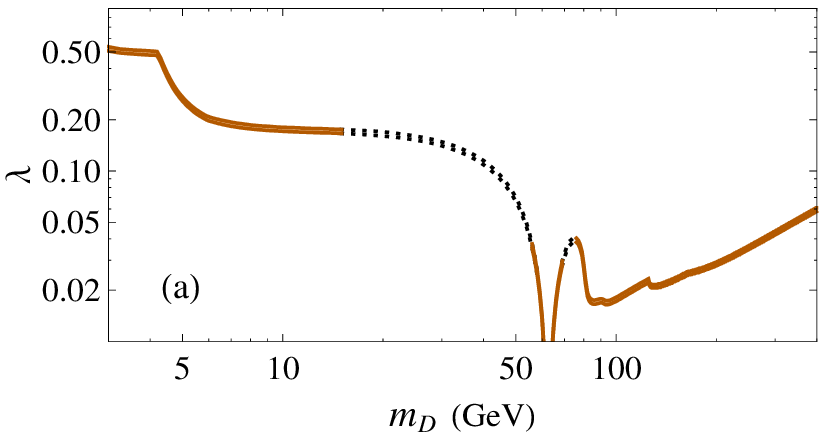}\vspace*{-1ex}\\
\includegraphics[width=83mm]{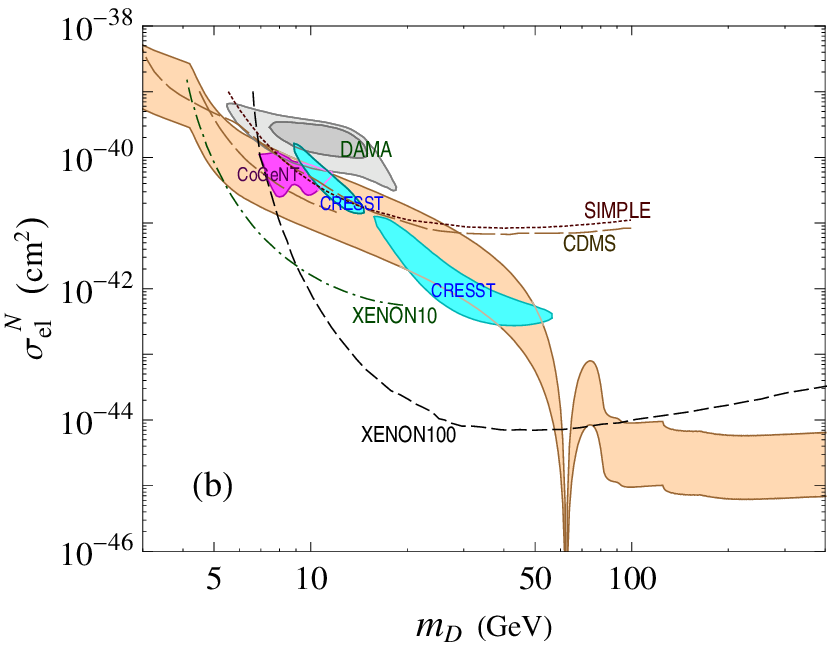} \, \,\,
\caption{(a)~Darkon-Higgs coupling $\lambda$ as a function of darkon mass $m_D^{}$ for
Higgs mass $m_h^{}=125$\,GeV  in~SM3+D.
(b)~The corresponding darkon-nucleon cross-section~$\sigma_{\rm el}^{N}$, compared~to
{90\%-CL}~upper limits from CDMS (brown long-dashed curves)~\cite{Akerib:2010pv},
XENON10 (green dot-dashed curve)~\cite{Angle:2011th},
XENON100 (black short-dashed curve)~\cite{Aprile:2011hi}, and
Stage 2 of SIMPLE (dark-red dotted curve)~\cite{Felizardo:2011uw}, as well as
the {90\%-CL} (magenta) signal region suggested by CoGeNT~\cite{Aalseth:2010vx},
a~gray (lighter gray) region compatible with the DAMA modulation signal at
the 3$\sigma$ (5$\sigma$) level~\cite{Bernabei:2010mq,Savage:2008er}, and two
2$\sigma$-confidence (cyan) areas representing the CRESST-II result~\cite{cresst}.
The black-dotted parts of the curve~in~(a) are disallowed by the direct-search limits
in~(b), after Ref.\,\cite{He:2011de}.\label{lambda_sm+d}}
\end{figure}

With the appearance of possible clues of a SM-like Higgs in the latest LHC
data~\cite{lhc}, we consider in this section some of the implications for the SM3+D
and~SM4+D.\footnote{\baselineskip=11pt%
Some other aspects of the darkon model or its close variants and its potential impact on
Higgs searches were treated before in~Ref.\,\cite{others}.
Alternative scenarios involving scalar dark matter which may also affect the Higgs sector
were dealt with recently in Ref.\,\cite{Englert:2011yb}.}
We follow Ref.\,\cite{He:2011de} to apply the procedure given in~Ref.\,\cite{He:2010nt} for
\,$3{\rm\,GeV}\le m_D^{}\le400$\,GeV,\, but now with the specific
selection~\,$m_h^{}=125$\,GeV\, for definiteness, in order to extract the darkon-Higgs coupling
$\lambda$ from the measured DM relic density~\,$\Omega_D^{}h^2=0.1123\pm 0.0035$~\cite{wmap7}.
We present the results for the SM3+D in~Fig.\,\ref{lambda_sm+d}(a), where the band width
reflects the 90\% confidence-level (CL) range~\,$0.092\le\Omega_D^{}h^2\le0.118$\, and the black-dotted
sections are ruled out by direct-search limits.
The $\lambda$ values in the SM4+D (not drawn) are roughly similar and mostly somewhat lower than
their SM3+D counterparts, by no more than~{\footnotesize$\sim$}20\%~\cite{He:2010nt,He:2011de}.
In Fig.\,\ref{lambda_sm+d}(b) we show the darkon-nucleon elastic cross-section
$\sigma_{\rm el}^{N}$ computed using the parameter choices in~Fig.\,\ref{lambda_sm+d}(a).
The band width of the $\sigma_{\rm el}^{N}$ curve arises mainly from the sizable uncertainty
of the Higgs-nucleon coupling,~\,$0.0011\le g_{NNh}^{}\le0.0032$\,~\cite{He:2011de}, due to
its dependence on the pion-nucleon sigma term~$\sigma_{\pi N}^{}$ which is not
well-determined~\cite{Ellis:2008hf}.
Although the displayed $\sigma_{\rm el}^{N}$ at each $m_D^{}$ value now varies by up to an order
of magnitude, we get a more realistic picture of how the model confronts the latest data from
the leading direct-searches for~WIMP~DM, which are also shown in~Fig.\,\ref{lambda_sm+d}(b).
In the SM4+D the majority of the $\sigma_{\rm el}^{N}$ numbers are
\,{\small$\sim$\,}50\%\, higher than their SM3+D counterparts~\cite{He:2010nt}.

\begin{figure}[b]
\includegraphics[width=77mm]{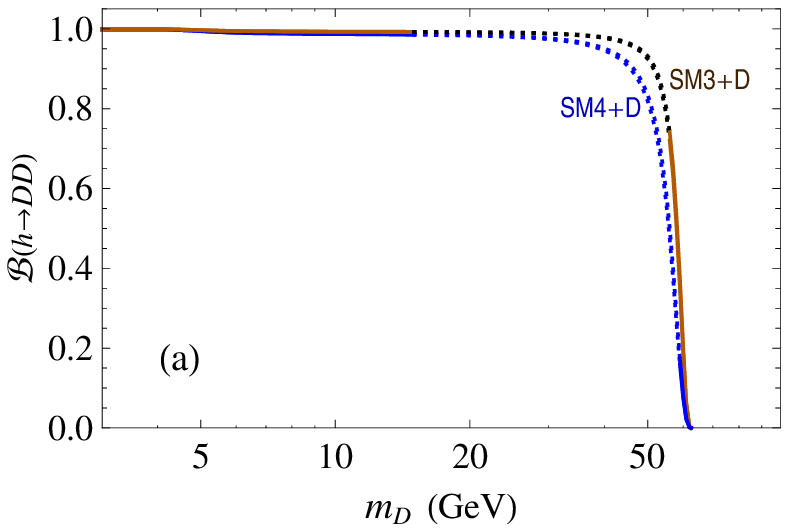} \, \,
\includegraphics[width=77mm,height=56mm]{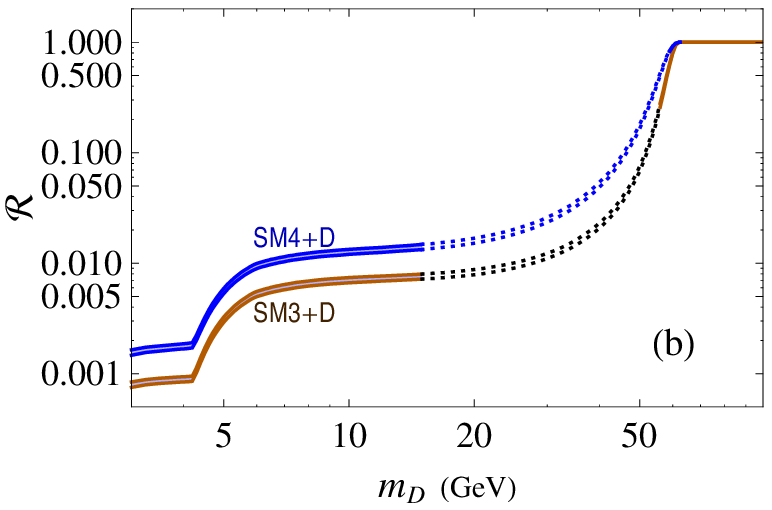}
\caption{(a) Branching ratio of \,$h\to DD$\, and (b)~the resulting reduction
factor~$\cal R$\, as functions of $m_D^{}$ in~SM3+D and~SM4+D
for \,$m_h^{}=125$~GeV.\,
The dotted parts are disallowed by direct-search limits.\label{br_sm+d}}
\end{figure}

If \,$m_h^{}>2m_D^{}$,\, the invisible decay channel \,$h\to DD$\, will become open with
branching ratio  \,${\cal B}(h\to DD)=\Gamma(h\to DD)/\Gamma_h^{\rm SM+D}$,\, where
\,$\Gamma(h\to DD)=\lambda^2v^2\bigl(1-4m_D^2/m_h^2\bigr){}^{1/2}/(8\pi m_h^{})$\, and
\,$\Gamma_h^{\rm SM+D}=\Gamma_h^{\rm SM}+\Gamma(h\to DD)$\, includes the Higgs total
width $\Gamma_h^{\rm SM}$ in the SM3 or SM4 without the darkon.
From the $\lambda$ values obtained above, we plot ${\cal B}(h\to DD)$
in~Fig.\,\ref{br_sm+d}(a), where the dotted portions are again excluded.
In~Fig.\,\ref{br_sm+d}(b) we display the corresponding reduction
factor~\cite{Burgess:2000yq,He:2011de}
\begin{eqnarray}
{\cal R} \,\,=\,\,
\frac{{\cal B}\bigl(h\to X\bar X\bigr)}{{\cal B}\bigl(h\to X\bar X\bigr)_{\rm SM}}
\,\,=\,\, \frac{\Gamma_h^{\rm SM}}{\Gamma_h^{\rm SM}+\Gamma(h\to DD)_{\rm SM+D}^{}}
\end{eqnarray}
which lowers all the SM3 (SM4) Higgs branching ratios in the~SM3+D (SM4+D) in the same way.
It is clear from these graphs that for \,$m_D^{}\le15$\,GeV\, the SM3+D Higgs would be
mainly invisible.
Consequently the observation of a 125-GeV SM3-like Higgs having nonnegligible branching
ratios in the visible channels would imply the exclusion of the SM3+D light-darkon region.
Moreover, the only surviving masses would be restricted to the vicinity of, but below,
the boundary~\,$m_D^{}=m_h^{}/2$.\,
A similar conclusion was drawn in Ref.\,\cite{Djouadi:2011md}.

In the SM4+D, the effect of a light darkon, via the reduction factor ${\cal R}$,
on the Higgs production rate may be ameliorated by the enhancement of the gluon-fusion
cross-section $\sigma(gg\to h)$ by up to {\small$\sim$}9 times due to the fourth-generation
quarks~\cite{Li:2010fu}.
To see explicitly whether this can leave some room for an SM4+D light darkon if the LHC sees
an SM3-like Higgs, we now compare with the corresponding rate in the SM3
without the darkon.
Thus for \,$m_h^{}=125$\,GeV,\, fourth-generation quark masses of order~500\,GeV,\,
and~\,$m_D^{}=15$\,GeV,\, we obtain the factor
\begin{eqnarray}
R_X^{} \,\,=\,\,
\frac{\sigma(pp\to h+{\rm anything})_{\rm SM4}^{}\,{\cal B}\bigl(h\to X\bar X\bigr)_{\rm SM4+D}}
{\sigma(pp\to h+{\rm anything})_{\rm SM3}^{}\,{\cal B}\bigl(h\to X\bar X\bigr)_{\rm SM3}}
\,\,\sim\,\, \frac{9~\Gamma_h^{\rm SM3}}{\Gamma_h^{\rm SM4}+\Gamma(h\to DD)_{\rm SM4+D}^{}}
\end{eqnarray}
for \,$X=\tau^-,c,b,W^{(*)}$, or $Z^{(*)}$\, in which case
\,$\Gamma\bigl(h\to X\bar X\bigr){}_{\rm SM4}^{}=
\Gamma\bigl(h\to X\bar X\bigr){}_{\rm SM3}^{}$,\,
but for \,$X=\gamma$\, there is extra suppression from
\,$\Gamma(h\to\gamma\gamma)_{\rm SM4}<\Gamma(h\to\gamma\gamma)_{\rm SM3}$\,
due to the new heavy fermions~\cite{Guo:2011ab}.
In view of~Fig.\,\ref{br_sm+d}(b), for \,$m_D^{}\le15$\,GeV\, we find
\begin{eqnarray}
R_X^{} \,\,<\,\, 0.09 ~.
\end{eqnarray}
We conclude that in the SM4+D with a light darkon the Higgs production event rates would not
be SM3-like and the light-darkon region would be ruled out by the detection of such a Higgs,
as in the SM3+D case.

In contrast, an SM3+D darkon with \,$m_D^{}\ge m_h^{}/2$\, would be in harmony with
the discovery of an SM3-like Higgs, as its decay pattern is not modified by
the darkon effect at leading order.
However, this mass region up to \,$m_D^{}\sim80$\,GeV\, is already forbidden by direct-search
limits, as Fig.\,\ref{lambda_sm+d}(b) indicates.
Darkon masses higher than {\small$\sim$}80\,GeV\, are still viable and will be probed by
future direct searches~\cite{He:2010nt}.
As for the SM4+D in this $m_D^{}$ region, the detection of such a~Higgs would also spell
trouble, being at odds with the SM4 prediction of considerably amplified Higgs production
cross-sections~\cite{Li:2010fu}.

\section{Two-Higgs-doublet model plus darkon\label{thdm+d}}

In this darkon model, the Higgs sector is the so-called type III of
the two-Higgs-doublet model~(THDM).
The general form of its Yukawa Lagrangian can be expressed as~\cite{thdm}
\begin{eqnarray} \label{LY}
{\cal L}_{\rm Y}^{} &=&
-\bar Q_{j,L}^{}\bigl(\lambda_1^{\cal U}\bigr)_{jl}\tilde H_1^{}{\cal U}_{l,R}^{}
- \bar Q_{j,L}^{}\bigl(\lambda_1^{\cal D}\bigr)_{jl} H_1^{} {\cal D}_{l,R}^{}
- \bar Q_{j,L}^{}\bigl(\lambda_2^{\cal U}\bigr)_{jl}\tilde H_2^{}{\cal U}_{l,R}^{}
- \bar Q_{j,L}^{}\bigl(\lambda_2^{\cal D}\bigr)_{jl} H_2^{} {\cal D}_{l,R}^{} \nonumber \\
&& -\;
\bar L_{j,L}^{}\bigl(\lambda_1^{E}\bigr)_{jl} H_1^{}E_{l,R}^{}
- \bar L_{j,L}^{}\bigl(\lambda_2^{E}\bigr)_{jl} H_2^{}E_{l,R}^{}
\;+\; {\rm H.c.} ~,
\end{eqnarray}
where summation over \,$j,l=1,2,3$\, is implied,
$Q_{j,L}^{}$ $\bigl(L_{l,L}^{}\bigr)$ denote the left-handed quark (lepton) doublets of
the three families, ${\cal U}_{l,R}^{}$ and ${\cal D}_{l,R}^{}$ $\bigl(E_{l,R}^{}\bigr)$ are
the right-handed quark (charged lepton) fields, $H_{1,2}^{}$~represent the Higgs doublets,
\,$\tilde H_{1,2}^{}=i\tau_2^{}H_{1,2}^*$,\, and hence $\lambda_{1,2}^{{\cal U,D},E}$ are
3$\times$3 matrices containing the Yukawa couplings.
In terms of the Higgs components,
\begin{eqnarray}
H_{a}^{} \,\,=\,\, \left(\begin{array}{c} h^+_{a} \vspace{1ex} \\ \frac{1}{\sqrt2}
\bigl(v_{a}^{}+ h_{a}^0 + i I_{a}^0\bigr) \end{array}\right ) \,,
\end{eqnarray}
where \,${a}=1,2$\, and $v_{a}^{}$ is the VEV of $H_{a}^{}$ satisfying \,$v_1^2+v_2^2=v^2$,\,
with~\,$v=246$\,GeV.\,
The fields $h^+_{a}$ and $I_{a}^{}$ can be expressed in terms of physical Higgs bosons $H^+$
and $A$ and the would-be Goldstone bosons $w$ and $z$ as
\begin{eqnarray}
\left(\begin{array}{c}h^+_1 \\ h^+_2\end{array}\right) &=&
\left(\begin{array}{rrr} \cos\beta && -\sin\beta \\ \sin\beta && \cos\beta \end{array}\right)
\left(\!\begin{array}{c}w^+\\H^+\end{array}\!\right) \,, \nonumber \\
\left(\begin{array}{c}I_1^{} \\ I_2^{} \end{array}\right) &=&
\left(\begin{array}{rrr}\cos\beta &&-\sin\beta \\ \sin\beta && \cos\beta\end{array}\right)
\left(\begin{array}{c}z\\A\end{array}\right) \,,
\end{eqnarray}
with  \,$\cos\beta=v_1^{}/v$\, and \,$\sin\beta=v_2^{}/v$,\, whereas $h_{1,2}^0$ are related
to the $CP$-even Higgs mass eigenstates $H$ and $h$ by
\begin{eqnarray}
\left(\begin{array}{c}h_1^0 \\ h_2^0 \end{array}\right) \,=\,
\left(\begin{array}{rrr} \cos\alpha && -\sin\alpha \\ \sin\alpha && \cos\alpha\end{array}\right)
\left(\begin{array}{c}H \\ h \end{array}\right) \,.
\end{eqnarray}

After the diagonalization of the fermion mass matrices
\,$M_{{\cal U,D},E}^{}=
\bigl(\lambda_1^{{\cal U,D},E}v_1^{}+\lambda_2^{{\cal U,D},E}v_2^{}\bigr)/\sqrt2$,\,
one can derive from ${\cal L}_{\rm Y}^{}$ the Lagrangian for the couplings of
$h_{1,2}^0$ to the fermions
\begin{eqnarray} \label{LY'}
{\cal L}_{\rm Y}' &=&
-\bar{\cal U}_L^{} \Biggl[
\Biggl(M_{\cal U}^{}-\frac{\lambda_2^{\cal U}v_2^{}}{\sqrt2}\Biggr)\frac{h_1^0}{v_1^{}} +
\Biggl(M_{\cal U}^{}-\frac{\lambda_1^{\cal U}v_1^{}}{\sqrt2}\Biggr)\frac{h_2^0}{v_2^{}}
\Biggr] {\cal U}_R^{}
\nonumber \\ && -\;
\bar{\cal D}_L^{} \Biggl[
\Biggl(M_{\cal D}^{}-\frac{\lambda_2^{\cal D}v_2^{}}{\sqrt2}\Biggr)\frac{h_1^0}{v_1^{}} +
\Biggl(M_{\cal D}^{}-\frac{\lambda_1^{\cal D}v_1^{}}{\sqrt2}\Biggr)\frac{h_2^0}{v_2^{}}
\Biggr] {\cal D}_R^{}
\nonumber \\ && -\;
\bar E_L^{} \Biggl[ \Biggl(M_E^{}-\frac{\lambda_2^Ev_2^{}}{\sqrt2}\Biggr)\frac{h_1^0}{v_1^{}} +
\Biggl(M_E^{}-\frac{\lambda_1^E v_1^{}}{\sqrt2}\Biggr)\frac{h_2^0}{v_2^{}} \Biggr]{\cal D}_R^{}
\;+\; {\rm H.c.} ~,
\end{eqnarray}
where now \,$M_{\cal U}^{}={\rm diag}\bigl(m_u^{},m_c^{},m_t^{}\bigr)$,\, etc., and
all the fermions in \,${\cal U}=(u~~c~~t)^{\rm T}$,\, etc., are mass eigenstates,
but \,$\lambda_{1,2}^{{\cal U,D},E}$\, in general are not also diagonal separately.
For each of the flavor-diagonal couplings in~${\cal L}_{\rm Y}'$, one can then write
in terms of the physical field \,${\cal H}=h$ or $H$
\begin{eqnarray}
{\cal L}_{ff\cal H}^{} \,\,=\,\, -k_f^{\cal H}\,m_f^{}\,\bar f f\,\frac{\cal H}{v} ~,
\end{eqnarray}
where for, say, the first family
\begin{eqnarray} & \displaystyle
k_u^h \,\,=\,\, \frac{\cos\alpha}{\sin\beta} \,-\,
\frac{\lambda_1^u\,v\,\cos(\alpha-\beta)}{\sqrt2\,m_u^{}\,\sin\beta} ~, \hspace{5ex}
k_u^H \,\,=\,\, \frac{\sin\alpha}{\sin\beta} \,-\,
\frac{\lambda_1^u\,v\,\sin(\alpha-\beta)}{\sqrt2\,m_u^{}\,\sin\beta} ~,
& \nonumber \\ & \displaystyle
k_d^h \,\,=\,\, -\frac{\sin\alpha}{\cos\beta} \,+\,
\frac{\lambda_2^d\,v\,\cos(\alpha-\beta)}{\sqrt2\,m_d^{}\,\cos\beta} ~, \hspace{5ex}
k_d^H \,\,=\,\, \frac{\cos\alpha}{\cos\beta} \,+\,
\frac{\lambda_2^d\,v\,\sin(\alpha-\beta)}{\sqrt2\,m_d^{}\,\cos\beta} ~,
& \nonumber \\ & \displaystyle
k_e^h \,\,=\,\, -\frac{\sin\alpha}{\cos\beta} \,+\,
\frac{\lambda_2^e\,v\,\cos(\alpha-\beta)}{\sqrt2\,m_e^{}\,\cos\beta} ~, \hspace{5ex}
k_e^H \,\,=\,\, \frac{\cos\alpha}{\cos\beta} \,+\,
\frac{\lambda_2^e\,v\,\sin(\alpha-\beta)}{\sqrt2\,m_e^{}\,\cos\beta} ~, \label{kf}
\end{eqnarray}
where \,$\lambda_{a}^{u,d,e}=\bigl(\lambda_{a}^{{\cal U,D},E}\bigr){}_{11}^{}$.\,
The corresponding $k_f^{\cal H}$ for the second and third families have analogous expressions.
Since only the combination \,$\lambda_1^f v_1^{}+\lambda_2^f v_2^{}=\sqrt2\,m_f^{}$\, is fixed
by the $f$ mass, $\lambda_{a}^f$~in $k_f^{\cal H}$ is a free parameter, and so is~$k_f^{\cal H}$.
We remark that setting \,$\lambda_1^{\cal U}=\lambda_2^{\cal D}=\lambda_2^E=0$\, leads to the type II of
the THDM+D studied in~Refs.\,\cite{He:2008qm,Cai:2011kb}.

Since the matrices \,$\lambda_{1,2}^{{\cal U,D},E}$\, in Eq.\,(\ref{LY'}) generally are
not diagonal, their off-diagonal elements may give rise to flavor-changing neutral
currents (FCNC) involving the Higgses at tree level.
We assume that these flavor-changing elements have their naturally small values
according to the Cheng-Sher {\it ansatz}~\cite{Cheng:1987rs}, namely,
\,$(\lambda_a)_{jl}^{}\sim \bigl(m_j^{}m_l^{}\bigr){}^{1/2}/v$\, for~\,$j\neq l$.\,
Since this {\it ansatz} is facing challenges from current experiments~\cite{Branco:2011iw},
we could suppress the FCNC effects further by increasing the mediating Higgs masses,
beyond which fine-tuning may be unavoidable.

Turning to the DM sector of the THDM+D, as in the SM+D, we ensure the darkon's stability as
a WIMP candidate by assuming $D$ to be a gauge singlet and introducing a~discrete $Z_2$
symmetry under which \,$D\to-D$,\, all the other fields being unaltered.
Its renormalizable Lagrangian then takes the form~\cite{Bird:2006jd,He:2007tt}
\begin{eqnarray} \label{LD2hdmd}
{\cal L}_D^{} \,\,=\,\, \mbox{$\frac{1}{2}$}\partial^\mu D\,\partial_\mu^{}D
-\mbox{$\frac{1}{4}$}\lambda_D^{}D^4 - \mbox{$\frac{1}{2}$}m_0^2D^2 \,-\,
\bigl[ \lambda_1^{}H_1^{\dag}H_1^{} + \lambda_2^{}H^\dagger_2 H_2^{} +
\lambda_3^{}\bigl(H_1^{\dag}H_2^{}+H_2^\dagger H_1^{}\bigr) \bigr] D^2 ~. ~~~~
\end{eqnarray}
The parameters in the potential of the model should be chosen such that $D$ does not
develop a~VEV and the $Z_2$ symmetry stays unbroken, so that $D$ does not mix with
the Higgs fields, maintaining its stability.
After electroweak symmetry breaking, Eq.\,(\ref{LD2hdmd}) contains the darkon mass
$m_D^{}$ and the $DD(h,H)$ terms \,$-\lambda_h^{}v\, D^2 h-\lambda_H^{}v\,D^2 H$,\,
but no $DDA$ coupling, where
\begin{eqnarray} \vphantom{|_{\displaystyle|_|}} & \displaystyle
m_D^2 \,\,=\,\, m_0^2 \,+\,
\bigl[\lambda_1^{}\cos^2\!\beta+\lambda_2^{}\sin^2\!\beta+\lambda_3^{}\sin(2\beta)\bigr]v^2 ~,
& \\ & \displaystyle
\lambda_h^{} \,\,=\,\, -\lambda_1^{}\sin\alpha\,\cos\beta+\lambda_2^{}\cos\alpha\,\sin\beta +
\lambda_3^{}\cos(\alpha+\beta) ~, & \nonumber \\ &
\lambda_H^{} \,\,=\,\,
\lambda_1^{}\cos\alpha\,\cos\beta+\lambda_2^{}\sin\alpha\,\sin\beta +
\lambda_3^{}\sin(\alpha+\beta) ~. & \label{lambda}
\end{eqnarray}
Since $m_0^{}$ and $\lambda_{1,2,3}^{}$ are free parameters, so are $m_D^{}$
and~$\lambda_{h,H}^{}$.
We note that for a heavy darkon with \,$m_D^{}>m_{h,H,A,H^+}^{}$\, the darkon annihilation
rate also gets contributions from $DD$ couplings to Higgs pairs
$\bigl(h^2,H^2,hH,AA,H^+H^-\bigr)$ which can be easily derived from Eq.\,(\ref{LD2hdmd}).

To evaluate the annihilation rates, the $h$ and $H$ couplings to the $W$ and $Z$ bosons
may be relevant depending on~$m_D^{}$.
The couplings are given by
\begin{eqnarray} \label{vvh}
{\cal L}_{VV\cal H}^{} \,\,=\,\, \frac{1}{v}
\bigl(2m_W^2\,W^{+\mu}W_\mu^-+m_Z^2\,Z^\mu Z_\mu^{}\bigr)
\bigl[h\,\sin(\beta-\alpha)+H\, \cos(\beta-\alpha)\bigr]
\end{eqnarray}
from the Higgs kinetic sector of the model~\cite{thdm}.

Inspired by the possible evidence for a 125-GeV SM-like Higgs in the LHC data,
we adopt
\begin{eqnarray} \label{cos}
\cos(\beta-\alpha) \,\,=\,\, 0 ~.
\end{eqnarray}
Applying one of its solutions, \,$\beta-\alpha=\pi/2$,\, to Eqs.\,(\ref{kf}), (\ref{lambda}),
and~(\ref{vvh}) yields
\begin{eqnarray}  & \displaystyle
k_u^h \,\,=\,\, k_d^h \,\,=\,\, k_e^h \,\,=\,\, 1 ~, &
\\ \nonumber \\ & \displaystyle
k_u^H \,\,=\,\, -\cot\beta\,+\,\frac{\lambda_1^u\,v}{\sqrt2\,m_u^{}\,\sin\beta} ~, \hspace{5ex}
k_d^H \,\,=\,\, \tan\beta \,-\, \frac{\lambda_2^d\,v}{\sqrt2\,m_d^{}\,\cos\beta} ~, &
\nonumber \\ \label{kH} & \displaystyle
k_e^H \,\,=\,\, \tan\beta \,-\, \frac{\lambda_2^e\,v}{\sqrt2\,m_e^{}\,\cos\beta} ~, ~~~~~~~ &
\\ \nonumber \\ \label{lambdahH} & \displaystyle
\lambda_h^{} \,\,=\,\,
\lambda_1^{}\cos^2\beta+\lambda_2^{}\sin^2\beta+\lambda_3^{}\sin(2\beta) ~, \hspace{5ex}
\lambda_H^{} \,\,=\,\, \mbox{$\frac{1}{2}$}\bigl(\lambda_1^{}-\lambda_2^{}\bigr)\sin(2\beta)
- \lambda_3^{}\cos(2\beta) ~, ~~~~~~~ &
\\ \nonumber \\ & \displaystyle
{\cal L}_{VV\cal H}^{} \,\,=\,\,
\bigl(2m_W^2\,W^{+\mu}W_\mu^-+m_Z^2\,Z^\mu Z_\mu^{}\bigr)\frac{h}{v} ~. &
\end{eqnarray}
Another consequence is that the tree-level flavor-changing couplings of $h$ vanish.
Evidently, now the couplings of $h$ to the SM fermions and gauge bosons are identical to those
of the SM Higgs.
The alternative solution, \,$\beta-\alpha=-\pi/2$,\, would yield the same results,
but with the opposite signs.
It is worth remarking that Eq.\,(\ref{cos}) is part of the \,$0\le|\cos(\beta-\alpha)|\ll1$\,
region of the model parameter space where $h$ can have SM-like couplings to the fermions and gauge
bosons, provided that $\tan\beta$ and $\cot\beta$, as well as the Higgs self-couplings,
are not large and that the $A$ mass is not below the electroweak scale~\cite{Gunion:2002zf}.

To render $h$ more SM-like, we require that the $hDD$ coupling
\,$\lambda_h^{}=0$.\footnote{\baselineskip=11pt%
The value of $\lambda_h$ may deviate somewhat from zero if $h$ is found with an invisible decay
rate that exceeds its SM range.
Since $\lambda_{1,2,3}$ in Eq.\,(\ref{lambdahH}) are free
parameters, the determination of $\lambda_h$ does not fix~$\lambda_H$.}
It follows that  for \,$m_D^{}<m_h^{}<m_H^{}$\,  the darkon annihilation contribution to
the DM relic density comes only from $H$-mediated diagrams.
Since $\lambda_1^u$ and $\lambda_2^{d,e}$ in Eq.\,(\ref{kH}) are free parameters,
for illustration we will pick for definiteness
\begin{eqnarray} \label{k=1}
k_u^H \,\,=\,\, k_d^H \,\,=\,\, k_e^H \,\,=\,\, 1 ~,
\end{eqnarray}
and similarly for $k_f^H$ belonging to the second and third families.
With these specific selections, $H$ share with $h$ the same couplings to the fermions, but
$H$ does not couple to the $W$ and $Z$ bosons at tree level, unlike~$h$.

\begin{figure}[b]
\includegraphics[width=70mm]{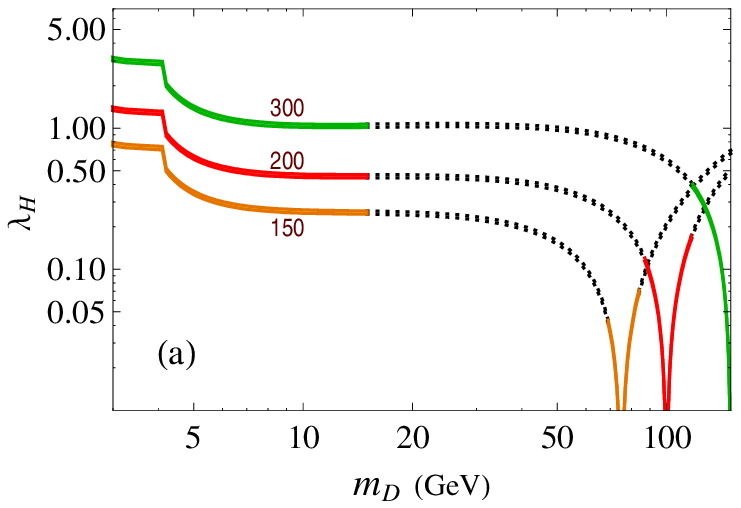} \, \, \,
\includegraphics[width=70mm]{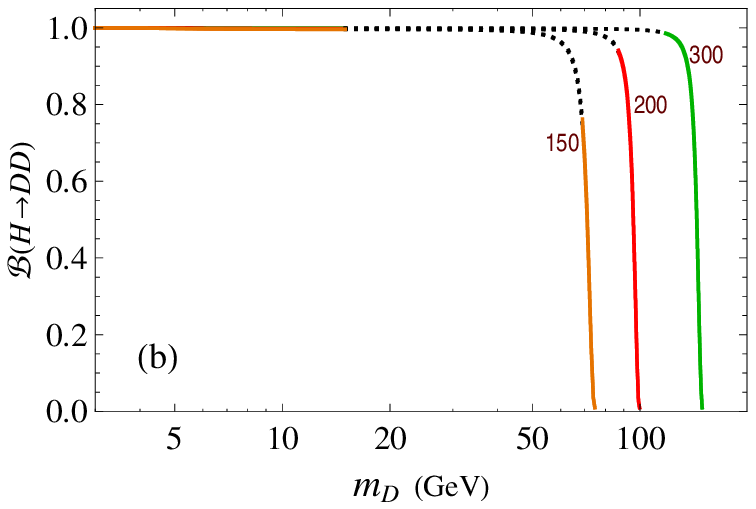}\\
\includegraphics[width=81mm]{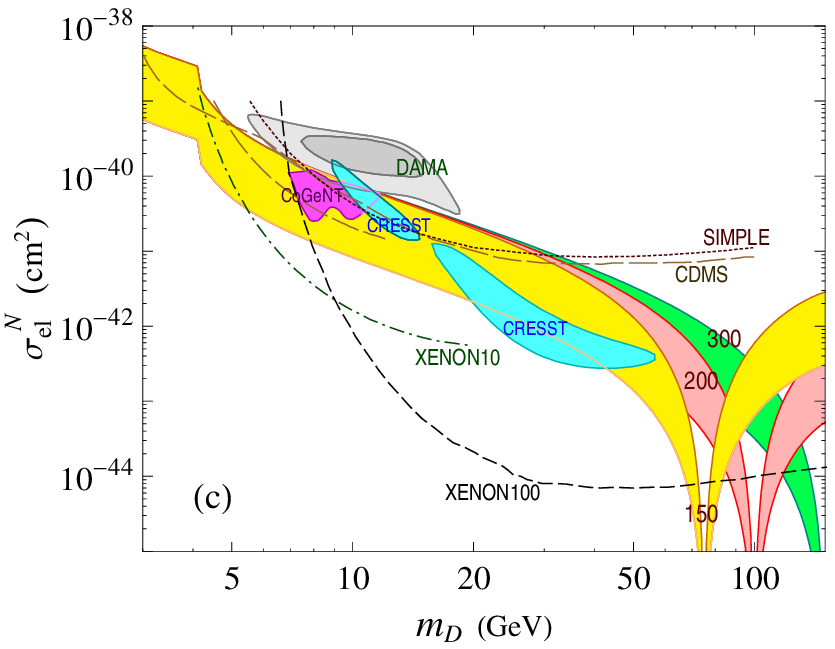} \, \,
\caption{(a)~Darkon-$H$ coupling $\lambda_H^{}$ as a function of darkon mass $m_D^{}$ for
\,$m_H^{}=150,200,300$~GeV,\, with the other couplings specified in the text, in the THDM+D
with isospin-conserving darkon-nucleon interactions.
The resulting (b) branching ratio of \,$H\to DD$\, and (c)~darkon-nucleon
cross-section~$\sigma_{\rm el}^{N}$, compared~to the same experimental results as
in~Fig.\,\ref{lambda_sm+d}(b).  The black-dotted parts in (a) and~(b) are disallowed
by direct-search bounds.\label{lambda_2hdm+d}}
\end{figure}

Since we are here interested in the case of a relatively light darkon, we concentrate on
the \,$m_D^{}\le150$\,GeV\, region.
Upon specifying~$m_H^{}$, one can extract $\lambda_H^{}$ from the relic-density data,
the procedure being similar to that in the preceding SM+D case.
We present the results for some illustrative values of $m_H^{}$ in~Fig.\,\ref{lambda_2hdm+d}(a),
where as before the dotted regions are forbidden by direct-search data.
From now on, we assume \,$m_D^{}\le m_H^{}<m_{A,H^\pm}^{}$.\,

The extracted $\lambda_H^{}$ translates into the branching ratio of invisible decay
\,$H\to DD$\, in~Fig.\,\ref{lambda_2hdm+d}(b).
As expected in the absence of $H(WW,ZZ)$ couplings at tree level, ${\cal B}(H\to DD)$ stays
close to~1 over most of the kinematically permitted range.
It is different from ${\cal B}(h\to DD)$ in the SM+D which becomes significantly less
dominant if \,$m_h^{}>2m_{W,Z}^{}$\,
after the important \,$h\to WW,ZZ$\, channels are open~\cite{He:2010nt,He:2011de}.

The cross section of elastic scattering of a darkon off a nucleon $N$ mediated by $H$
is~\cite{He:2008qm}
\begin{eqnarray}
\sigma_{\rm el}^N \,\,=\,\,
\frac{\lambda_H^2\,g_{NNH}^2\,v^2\,m_N^2}{\pi\,\bigl(m_D^{}+m_N^{}\bigr)^2\, m_H^4} \,\,,
\end{eqnarray}
where $g_{NNH}$ is the effective $H$-nucleon coupling.
With the choices \,$k_f^H=1$,\, as in Eq.\,(\ref{k=1}), which conserve isospin, $g_{NNH}$ is
no different from $g_{NNh}^{}$ used in the previous section.
Hence we employ~\,$0.0011\le g_{NNH}^{}\le0.0032$.\,

We show in Fig.\,\ref{lambda_2hdm+d}(c) the calculated $\sigma_{\rm el}^{N}$ corresponding to
the parameter selections in~Fig.\,\ref{lambda_2hdm+d}(a).
Also on display are the results of recent DM direct searches, as in~Fig.\,\ref{lambda_sm+d}(b).
One can see that, much like the SM+D case, the THDM+D prediction for the darkon-nucleon
cross-section can accommodate well especially the light-WIMP regions suggested by
CoGeNT and CRESST-II data~\cite{Aalseth:2010vx,cresst}, although they are in tension
with the null results of other direct searches.
However, unlike the~SM+D, the THDM+D still has enough parameter space to allow the particular
choices we made which yield both a SM-like Higgs, $h$, and a DM sector possessing a viable light
WIMP coupled to another Higgs,~$H$.

\section{Isospin-violating dark matter in THDM+D\label{ivdm}}

Up to now, we have only achieved making the THDM+D have a low-mass WIMP candidate and a SM-like
Higgs boson as hinted at by the LHC findings.
If the model is also to accommodate both DM direct-search results which indicated light-WIMP
evidence and those which did not, it needs to have a mechanism that can provide substantial
isospin violation in the WIMP interactions with nucleons.
As was proposed in the literature~\cite{Kurylov:2003ra,Feng:2011vu,Kopp:2011yr}, the tension in
the light-WIMP data will partially go away if the WIMP effective couplings $f_{p,n}^{}$ to
the proton and neutron, respectively, satisfy the relation~\,$f_n^{}\sim-0.7f_p^{}$.\,
The THDM+D may be able to realize this using the freedom still available in
the parameters~$k_f^H$ defined above.
By allowing them to deviate from the choices \,$k_f^H=1$\, in the last section,
which respect isospin, it may be feasible for the model to attain the desired results.
We explore this scenario in the following.

The WIMP-nucleon cross-section $\sigma_{\rm el}^N$ in the isospin-symmetric limit can be
expressed in terms of the WIMP-proton elastic cross-section $\sigma_{\rm el}^p$ in the presence
of isospin violation as~\cite{Kurylov:2003ra,Feng:2011vu}
\begin{eqnarray}
\sigma_{\rm el}^{N}\,f_p^2\sum_i\eta_i^{}\,\mu_{A_i}^2\,A_i^2 \,\,=\,\,
\sigma_{\rm el}^p\sum_i\eta_i^{}\,\mu_{A_i}^2
\bigl[{\cal Z}f_p^{}+\big(A_i^{}-{\cal Z}\bigr)f_n^{}\bigr]^2 ~,
\end{eqnarray}
where the sum is over the isotopes of the element in the detector material with which the WIMP
interacts dominantly,  $\cal Z$ is proton number of the element, $A_i$ $(\eta_i^{})$ each denote
the nucleon number (fractional abundance) of its isotopes,
and~\,$\mu_{A_i}=m_{A_i}m_{\rm WIMP}^{}/\bigl(m_{A_i}+m_{\rm WIMP}^{}\bigr)$\, involving
the isotope and WIMP masses.
Thus if isospin violation is negligible, \,$f_n^{}=f_p^{}$,\, the measurement of event rates of
WIMP-nucleus scattering will translate into the usual~\,$\sigma_{\rm el}^{N}=\sigma_{\rm el}^p$.\,
For~\,$f_n^{}=-0.7f_p^{}$,\, taking into account the different $A_i$ and $\cal Z$ numbers for the
different detector materials, one can transform some of the contradictory data on the WIMP-nucleon
cross-sections into $\sigma_{\rm el}^p$ numbers which overlap with each
other~\cite{Feng:2011vu,Kopp:2011yr}.
This also makes the extracted $\sigma_{\rm el}^p$ enhanced relative to the current measured
values of $\sigma_{\rm el}^{N}$ by up to 4 orders of magnitude, depending on $A_i$ and~$\cal Z$.

Now, in the THDM+D with only $H$ mediating the WIMP-nucleon interactions
\begin{eqnarray} \label{csel}
\sigma_{\rm el}^p \,\,=\,\, \frac{4\,m_D^2\,m_p^2\,f_p^2}{\pi\bigl(m_D^{}+m_p^{}\bigr)^2} ~,
\hspace{5ex} f_p^{} \,\,=\,\, \frac{\lambda_H^{}\,g_{ppH}^{}\,v}{2\,m_D^{}\,m_H^2} ~,
\end{eqnarray}
where the $H$-proton effective coupling $g_{ppH}^{}$ contains various quark contributions
according to~Eq.\,(\ref{gNNH}).
The relation \,$f_n^{}=\rho\,f_p^{}$\, then implies
\begin{eqnarray} \label{n=rp}
g_{nnH}^{} \,\,=\,\, \rho\,g_{ppH}^{} ~,
\end{eqnarray}
where the $H$-neutron effective coupling $g_{nnH}^{}$ also has the general form
in~Eq.\,(\ref{gNNH}) and we will set~\,$\rho=-0.7$.\,
Also relevant is the darkon annihilation-rate formula~\cite{Burgess:2000yq,He:2008qm}
\begin{eqnarray} \label{csan}
\sigma_{\rm ann}^{} v_{\rm rel}^{} \,\,=\,\,
\frac{4\lambda_H^2 v^2}{\bigl(4m_D^2-m_H^2\bigr)^2+\Gamma^2_H\,m^2_H}\,
\frac{\sum_i\Gamma\bigl(\tilde H\to X_i\bigr)}{m_D} \,\,,
\end{eqnarray}
where  $v_{\rm rel}^{}$\, is the relative speed of the $DD$ pair in their center-of-mass frame,
$\tilde H$  is a~virtual $H$ having the same couplings to other states as the physical $H$ of
mass $m_H^{}$, but with an invariant mass \,$\sqrt s=2m_D^{}$, and \,$\tilde H\to X_i$\, is
any kinematically permitted decay mode of $\tilde H$.

Using Eqs.\,(\ref{csel})-(\ref{csan}) and focusing on the low-mass
range~\,$5{\rm\,GeV}\le m_D^{}\le20$\,GeV,\, we scan the parameter space of the products
\,$\lambda_H^{}k_f^H$,\, as the factors always go together in Eqs.~(\ref{csel}) and~(\ref{csan}),
while imposing \,$\rho=-0.7$\, and the relic-density constraint.
We find that to enhance $\sigma_{\rm el}^p$ by a few orders of magnitude under these restrictions
implies that $k_{u,d}^{H}$ have to be big, \,$k_u^{H}\sim-2k_d^{H}$,\, and the other $k_f^H$
become negligible by comparison, confirming the finding of Ref.\,\cite{Gao:2011ka}.
For instance, with~\,$m_H^{}=200\;(300)$\,GeV\, we obtain
\,$0.6\;(1.4)\times10^3\le\lambda_H^{}k_u^H\le 0.8\;(1.8)\times10^3$\, corresponding
to~\,$5{\rm\,GeV}\le m_D^{}\le20$\,GeV.\,
It follows that in general \,$k_u^H={\cal O}\bigl(10^3\bigr)$\, if
\,$\lambda_H^{}={\cal O}(1)$\, and $m_H^{}$ is a few hundred~GeV.
For such large $k_{u,d}^{H}$, one expects that \,$k_u^{H}\sim\lambda_1^u v_1^{}/m_u^{}$\, and
\,$k_d^H\sim\lambda_2^d v_2^{}/m_d^{}$\,  from~Eq.\,(\ref{kH}).
Consequently, since \,$\lambda_1^u v_1^{}+\lambda_2^u v_2^{}=\sqrt2\,m_u^{}$\, and
\,$\lambda_1^d v_1^{}+\lambda_2^d v_2^{}=\sqrt2\,m_d^{}$,\,  some degree of subtle cancelations
between the \,$\lambda_a^{u,d}v_a^{}$\, terms is needed to reproduce the small $u$ and $d$ masses.
This is the price one has to pay for the greatly amplified~$\sigma_{\rm el}^p$.

\begin{figure}[t]
\includegraphics[width=123mm]{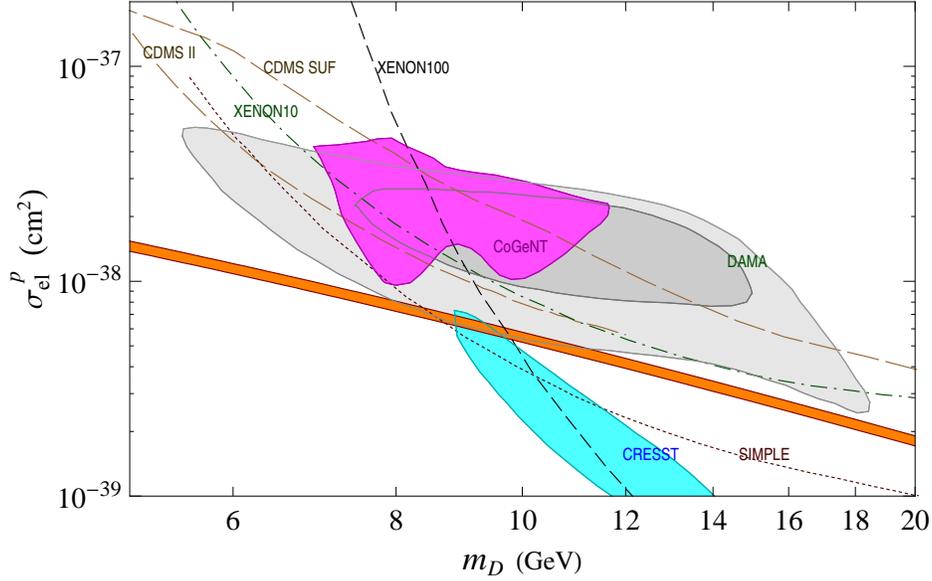}
\caption{Darkon-proton cross-section $\sigma_{\rm el}^p$ in THDM+D with
isospin-violating light darkon (orange curve) compared with several
direct-search results for WIMP-nucleon couplings
satisfying~\,$f_n^{}=-0.7f_p^{}$.\label{cs-ivdm}}
\end{figure}

We plot the theoretical cross-section (orange curve) in~Fig.\,\ref{cs-ivdm}, where
the band width reflects the relic-density uncertainty.  Also plotted are the direct-search
results~\cite{Akerib:2010pv,Angle:2011th,Aprile:2011hi,Felizardo:2011uw,Aalseth:2010vx,
Bernabei:2010mq,cresst} reproduced with the WIMP-nucleon couplings
satisfying~\,$f_n^{}=-0.7f_p^{}$.\,
In this $m_D^{}$ range, the theory curve is roughly independent of
$m_H^{}$ if it is in the hundreds of~GeV.
One can see that the prediction is not able to reach the (gray) region implied by
DAMA at the~3$\sigma$ level and only marginally covers its $5\sigma$ (lighter gray) region.
The CoGeNT preferred (magenta) area is also unreachable.  Nevertheless, with respect to
many points in these regions, the prediction is too low by no more than a factor of~2~or~3.
On the other hand, with appropriately lowered $k_{u,d}^H$, it can agree well with
the cross sections (cyan patch) favored by the CRESST-II results.
Furthermore, for $m_D^{}$ values below 10\,GeV or so the prediction curve does not conflict
with the XENON limits, which is unlike the isospin-symmetric case
illustrated in Fig.\,\ref{lambda_2hdm+d}.
The situation is very different when it comes to the limit from SIMPLE which now rules out
\,$m_D^{}\,\mbox{\footnotesize$\gtrsim$}\,9$\,GeV,\, while previously it allowed most of
the darkon parameter space.
On the experimental side, due to~\,$f_n^{}=-0.7f_p^{}$,\, virtually none of the CRESST-II
area in~Fig.\,\ref{lambda_2hdm+d} is consistent with the DAMA and CoGeNT ones any more,
and the SIMPLE bound disallows most of the CoGeNT area.
It is obvious from a~comparison of Figs.~\ref{lambda_2hdm+d}(c) and~\ref{cs-ivdm} that there are
unresolved puzzles remaining.
To address them in a comprehensive manner would likely have to await future direct-searches with
improved precision and may need to involve extra ingredients~\cite{Kopp:2011yr} beyond the simple
frameworks treated in this work.

We now demonstrate that the result above for the enhanced $\sigma_{\rm el}^p$ prediction does
not depend on the value of $k_u^H$ or the pion-nucleon sigma term~$\sigma_{\pi N}^{}$, assuming
that $k_{u,d}^H$ are much bigger than the other~$k_f^H$.
As discussed in~Appendix\,\ref{eff}, the requirements \,$k_{u,d}^H\gg k_{s,c,b,t}^{H}\sim0$\,
and~\,$g_{nnH}^{}=\rho\,g_{ppH}^{}$\, result in
\begin{eqnarray}
g_{ppH}^{} \,\,=\,\,
\frac{3.0\times10^{-5}~k_u^{H}\,\sigma_{\pi N}^{}}
{5.2(1+\rho){\rm\,MeV}\,+\,(1-\rho)\sigma_{\pi N}^{}}
\end{eqnarray}
and also \,$k_d^H\propto k_u^H\,$ according to~Eq.\,(\ref{kdH}).
Hence for \,$-1\lesssim\rho\lesssim-0.5$\, and~\cite{He:2011de}
\,$\sigma_{\pi N}^{}\ge30$\,MeV\,
\begin{eqnarray}
g_{ppH}^{} \,\,\sim\,\, \frac{3\times10^{-5}~k_u^{H}}{(1-\rho)} ~, \hspace{5ex}
k_d^H \,\,\sim\,\, -0.5\,k_u^H
\end{eqnarray}
approximately independent of $\sigma_{\pi N}^{}$.
Since now $g_{ppH}^2$ and
\,$\Sigma_i^{}\Gamma_{\tilde H\to X_i}\simeq
\Gamma_{\tilde H\to u\bar u}+\Gamma_{\tilde H\to d\bar d}$\,
are both roughly proportional to $\bigl(k_u^{H}\bigr){}^2$,
the approximate expression
\begin{eqnarray}
\sigma_{\rm el}^p \,\,\simeq\,\,
\frac{g_{ppH}^2\,m_D^{}\,m_p^2}{4\pi\,\bigl(m_D^{}+m_p^{}\bigr)^2}~
\frac{\sigma_{\rm ann}^{}v_{\rm rel}^{}}
{\Gamma_{\tilde H\to u\bar u}+\Gamma_{\tilde H\to d\bar d}} ~,
\end{eqnarray}
derived from Eqs.~(\ref{csel}) and~(\ref{csan}) and valid for \,$m_p^{}<m_D^{}\ll m_H^{}$,\,
is roughly independent of $k_u^{H}$.
Consequently, $\sigma_{\rm el}^p$ can only be increased further if
\,$\sigma_{\rm ann}^{}v_{\rm rel}^{}$\, is also increased.

Finally, it is worth mentioning that one cannot obtain the substantial isospin violation of interest
if the Higgs sector of the THDM+D is of type II, in which the up- and down-type quarks are
coupled to different Higgs doublets~\cite{thdm}.
In that case, the second term in each of the formulas for $k_f^H$ in Eq.\,(\ref{kH})
is absent, and therefore there is not much freedom to vary~$k_f^H$, as $\tan\beta$ cannot be
arbitrarily small or big due to restrictions from various data~\cite{gfitter1,gfitter2}.
Accordingly, since the darkon-nucleon couplings are dominated by the combined
strange- and heavy-quark contributions, which conserve isospin, as can be seen from
Eqs.~(\ref{gNNH}) and~(\ref{fq}), the darkon-proton coupling cannot be made large enough after
applying the relic density and \,$g_{nnH}^{}=-0.7\,g_{ppH}^{}$\, restraints.
We have found that the resulting darkon-proton cross-section $\sigma_{\rm el}^p$ for
\,$m_D^{}\sim10$\,GeV\, cannot reach more than \,{\footnotesize$\sim$}$10^{-43}{\rm\,cm}^2$,\,
about 5 orders of magnitude too small.

\section{Conclusions\label{concl}}

The preliminary indications of a Higgs boson with SM-like properties in the latest LHC data, if
confirmed by future measurements, will have important implications for WIMP dark matter models.
We have explored a number of such implications for some of the simplest darkon models.
For the simplest one, SM3+D, most of the light-darkon mass range will be ruled out
if an SM3-like Higgs with mass near 125\,GeV is found.
Such a discovery would also be at odds with the SM4 Higgs expectations, thus disfavoring the SM4+D.
In contrast, the type-III two-Higgs-doublet model enlarged with the addition of a darkon has
an abundance of allowed parameter space in its DM sector.
It can accommodate an~SM3-like Higgs and simultaneously offers a WIMP candidate in harmony with
the light-WIMP hypothesis inspired by the clues from a number of DM searches,
although it is in conflict with the null results of other searches.
The model can also provide substantial isospin violation in the WIMP-nucleon interactions
which can alleviate some of this tension in the data on light DM.
However, this could be achieved only with some amount of fine-tuning in several of
the relevant parameters.
Nevertheless, upcoming searches for the Higgs at the LHC and future DM direct searches can
test further the THDM+D which we have considered.
Additional signals of the darkon in this scenario may be available from flavor-changing
top-quark decays into lighter up-type quarks with missing energy which are potentially
observable at the~LHC.

\acknowledgments

This work was supported in part by NSC of ROC, NNSF and SJTU 985 grants of PRC, and
NCU Plan to Develop First-Class Universities and Top-Level Research Centers.

\appendix

\section{Higgs-nucleon effective couplings\label{eff}}

The effective coupling of a Higgs $\cal H$ to a proton $p$ or neutron $n$ is related to
the quark Yukawa parameters by~\cite{He:2008qm,Shifman:1978zn}
\begin{eqnarray} \label{gNNH}
g_{\cal NNH}^{}\, \bar u_{\cal N}^{}u_{\cal N}^{} \,\,=\,\,
\bar u_{\cal N}^{}u_{\cal N}^{} \sum_q g_q^{\cal N} k_q^{\cal H} \,\,=\,\,
\sum_q\frac{k_q^{\cal H}}{v}\langle{\cal N}|m_q^{}\,\bar q q|{\cal N}\rangle ~, \hspace{5ex}
{\cal N} \,\,=\,\, p,n ~,
\end{eqnarray}
where $u_{\cal N}^{}$ is the Dirac spinor for $\cal N$ and the sum is over all quarks.
Using the chiral Lagrangian approach described in Ref.\,\cite{He:2008qm}, but without
neglecting isospin violation, and assuming three fermion families, we obtain
\begin{eqnarray} & \displaystyle
g_u^p \,\,=\,\, \frac{-2\bigl(b_D^{}+b_F^{}+b_0^{}\bigr)m_u^{}}{v} ~, \hspace{5ex}
g_u^n \,\,=\,\, \frac{-2 b_0^{}\,m_u^{}}{v} ~, & \\ & \displaystyle
g_d^p \,\,=\,\, \frac{-2 b_0^{}\,m_d^{}}{v} ~, \hspace{5ex}
g_d^n \,\,=\,\, \frac{-2\bigl(b_D^{}+b_F^{}+b_0^{}\bigr)m_d^{}}{v} ~, & \\ & \displaystyle
g_s^p \,\,=\,\, g_s^n \,\,=\,\, \frac{-2\bigl(b_D^{}-b_F^{}+b_0^{}\bigr)m_s^{}}{v} ~, \hspace{5ex}
g_{c,b,t}^p \,\,=\,\, g_{c,b,t}^n \,\,=\,\, \frac{2\,m_B^{}}{27\,v} ~, & \\ & \displaystyle
\sigma_{\pi N}^{} \,\,=\,\, -\bigl(b_D^{}+b_F^{}+2b_0^{}\bigr)\bigl(m_u^{}+m_d^{}\bigr) ~, &
\end{eqnarray}
where the parameters $b_{D,F,0}^{}$ and $m_B^{}$ can be fixed from the measured masses of
the lightest baryons and the phenomenological or lattice value of the pion-nucleon sigma
term~$\sigma_{\pi N}^{}$.
Since $\sigma_{\pi N}^{}$ is still poorly determined~\cite{Ellis:2008hf}, we take
\,$30{\rm\,MeV}\le\sigma_{\pi N}^{}\le80$\,MeV\, after~Ref.\,\cite{He:2011de}.
Hence for \,$\sigma_{\pi N}^{}=30~(80)~$MeV\, the values of $g_q^{\cal N}$ are,
in units of $10^{-3}$,
\begin{eqnarray} &
g_u^p \,\,=\,\, 0.05~(0.12) ~, \hspace{5ex} g_u^n \,\,=\,\, 0.04~(0.11) ~,
& \nonumber \\ &
g_d^p \,\,=\,\, 0.06~(0.20) ~, \hspace{5ex} g_d^n \,\,=\,\, 0.09~(0.22) ~,
& \nonumber \\ \label{fq} &
g_s^{p,n} \,\,=\,\, 0.25~(2.88) ~, \hspace{5ex} g_{c,b,t}^{p,n} \,\,=\,\, 0.26~(0.05). ~ &
\end{eqnarray}

In the following we assume that \,$k_{u,d}^H\neq0$,\, the other~$k_f^H$ are zero, and
\,$g_{nnH}^{}=\rho\, g_{ppH}^{}$\, as in~Eq.\,(\ref{n=rp}), with $\rho$ being a~constant.
Combining these requirements with some of the preceding equations, we arrive at
\begin{eqnarray} & \displaystyle
k_d^H \,\,=\,\,
\frac{(1+\rho)\bigl(b_D^{}+b_F^{}\bigr)\bigl(m_u^{}+m_d^{}\bigr)+(1-\rho)\sigma_{\pi N}^{}}
{(1+\rho)\bigl(b_D^{}+b_F^{}\bigr)\bigl(m_u^{}+m_d^{}\bigr)-(1-\rho)\sigma_{\pi N}^{}}~
\frac{m_u^{}}{m_d^{}}\, k_u^H ~,
& \\ \nonumber \\ & \displaystyle
g_{ppH}^{} \,\,=\,\,
\frac{4\bigl(b_D^{}+b_F^{}\bigr)m_u^{}\,k_u^{H}\,\sigma_{\pi N}^{}/v}
{\bigl(b_D^{}+b_F^{}\bigr)\bigl(m_u^{}+m_d^{}\bigr)(1+\rho)-(1-\rho)\sigma_{\pi N}^{}} ~. &
\end{eqnarray}
Numerically we get
\,$\bigl(b_D^{}+b_F^{}\bigr)\bigl(m_u^{},m_d^{}\bigr)\simeq(-1.9,-3.3)$\,MeV,\,
which in this case leads to
\begin{eqnarray} \label{kdH} & \displaystyle
k_d^H \,\,\simeq\,\, \frac{2.9(1+\rho){\rm\,MeV}\,-\,0.56\,(1-\rho)\sigma_{\pi N}^{}}
{5.2(1+\rho){\rm\,MeV}\,+(1-\rho)\sigma_{\pi N}^{}}~ k_u^H ~,
& \\ \nonumber \\ \label{gppH} & \displaystyle
g_{ppH}^{} \,\,\simeq\,\, \frac{3.0\times10^{-5}~k_u^{H}\,\sigma_{\pi N}^{}}
{5.2(1+\rho){\rm\,MeV}\,+\,(1-\rho)\sigma_{\pi N}^{}} ~. &
\end{eqnarray}
In evaluating these quantities and \,$\Sigma_i^{}\Gamma_{\tilde H\to X_i}$\, in
Eq.\,(\ref{csan}),
we have employed the running masses of the light quarks at scales $\,\mu=1$\,GeV\, and
\,$\mu=2m_D^{}$,\, respectively, and included QCD corrections in the
\,$\tilde H\to q\bar q$\, rates~\cite{Djouadi:2005gi}.
The ratios of light-quark masses used are \,$m_d^{}/m_u^{}\simeq1.8$\, and
\,$m_s^{}/m_d^{}\simeq20$,\, which fall within their measured ranges~\cite{Nakamura:2010zzi}.

\end{document}